# Comparative study of Gamow-Teller strength distributions on odd-odd nucleus $^{50}$V and its impact on electron capture rates in astrophysical environments

**Jameel-Un Nabi**[*], **Muhammad Sajjad**

Faculty of Engineering Sciences, Ghulam Ishaq Khan Institute of Engineering
Sciences and Technology, Topi 23640, Swabi, NWFP, Pakistan

Gamow-Teller (GT) strength transitions are an ideal probe for testing nuclear structure models. In addition to nuclear structure, GT transitions in nuclei directly affect the early phases of Type Ia and Type-II supernovae core collapse since the electron capture rates are partly determined by these GT transitions. In astrophysics, GT transitions provide an important input for model calculations and element formation during the explosive phase of a massive star at the end of its life-time. Recent nucleosynthesis calculations show that odd-odd and odd-A nuclei cause the largest contribution in the rate of change of lepton-to-baryon ratio. In the present manuscript, we have calculated the GT strength distributions and electron capture rates for odd-odd nucleus $^{50}$V by using the pn-QRPA theory. At present $^{50}$V is the first experimentally available odd-odd nucleus in fp-shell nuclei. We also compare our GT strength distribution with the recently measured results of a $^{50}$V(d,$^2$He)$^{50}$Ti experiment, with the earlier work of Fuller, Fowler, and Newman (referred to as FFN) and subsequently with the large-scale shell model calculations. SNe Ia model calculations performed using FFN rates result in overproduction of $^{50}$Ti, and were brought to a much acceptable value by employing shell model results. It might be interesting to study how the composition of the ejecta using presently reported QRPA rates compare with the observed abundances.

**PACS** number(s): 26.20.+f, 21.60.Jz, 23.40.-s, 26.50.+x

## I. INTRODUCTION

The Gamow-Teller (GT) response is astrophysically important for a number of nuclides, particularly around iron. One of the factors controlling the late phases of stellar evolution and initiation of stellar ... is the lepton fraction [1,2]. This ... imarily beta-decay and electron

Corresponding author.
E-mail address: jnabi00@gmail.com
Fax: +92-938-271890

capture rates). An important and non-trivial contribution to the weak rates is the distribution of GT strength. In recent years, new developments in nuclear-structure physics have led to renewed interest in the role of the GT strength distribution in fp-shell nuclei. If the mass of the iron core exceeds the Chandrasekhar limit, the pressure of the relativistic degenerate electron gas can no longer resist the force of gravity, and the core starts to collapse marking the beginning of Type-II supernova. SNe Ia are thought to be the explosions of white dwarfs that accrete matter from binary companions. The relatively high Fermi energy of the degenerate electron gas in white dwarfs can lead to substantial electron capture thereby significantly reducing the lepton fraction. This then controls the isotopic composition ejected from such explosions. GT strengths have important implications in this scenario as well, such as explosive nucleosynthesis in O-Ne-Mg white dwarfs [e.g. 3, 4].

The first extensive effort to tabulate the weak interaction rates at high temperatures and densities, where decays from excited states of the parent nuclei become relevant, was done by Fuller, Fowler, and Newman (FFN) [2]. FFN calculated the weak interaction rates over a wide range of temperatures and densities ($10 \leq \rho Y_e$ (g cm$^{-3}$) $\leq 10^{11}$, $10^7 \leq$ T (K) $\leq 10^{11}$). The GT strength and excitation energies were calculated using a zero-order shell model. They also incorporated the experimental data available at that time. The matrix elements of Brown et al. [5] were used for unmeasured GT strengths. When these were also not available, FFN assumed an average log $ft$ value of 5.0 for their calculation of weak rates. Later Aufderheide et al. [6] extended the FFN work for neutron rich nuclei in pf-shell. They tabulated the top 90 electron capture nuclei averaged throughout the stellar trajectory (see Table 25 therein) for core collapse densities $\rho = (10^7 - 10^{10})$ g/cm$^3$. They found that for densities $\rho_7 > 10$ ($\rho_7$ measures the density in $10^7$ g/cm$^3$) electrons are most effectively captured by odd-odd nuclei. These were the nuclei which, according to the calculations of the authors, affected $\dot{Y}_e$ (where $Y_e$ is the lepton-to-baryon ratio) the most in the presupernova evolution. With the advent of high computing machines and efficient algorithms, large-scale shell-model calculations have also been performed for pf-shell nuclei in the mass range A= 45-65 [7] and employed in many simulation codes with relative success [8].

Recent nucleosynthesis calculations [e.g. 8] show that the rate of change of lepton fraction (product of rates and nuclear abundances) is significantly affected by odd-odd and odd-A nuclei, while the contribution of even-even nuclei is negligible. This also implies that the largest contribution to the neutronization of the ejecta comes from odd-odd and odd-A nuclei. The list of most important electron capture nuclei (those which make the largest contribution to $\dot{Y}_e$) compiled by Aufderheide and collaborators [6] contained the isotope of vanadium, $^{50}$V. The $^{50}$V nucleus is one of the very few stable odd-odd nuclei in the *pf* shell and is the only odd-odd nucleus experimentally accessible at present.

The GT strength distribution can be determined using the *(d,$^2$He)* reaction [9, 10]. High-resolution spectra for this reaction are now available for light nuclei [11] and this program is being extended to medium-weight nuclei. The available experimental data obtained for the GT strength distribution have until recently been hampered by a rather poor energy resolution, making a detailed comparison between theory and experiment difficult [12]. It is, however, important to note that the GT strength is fragmented over many discrete states in the *fp*-shell nuclei, and the details of the strength distributions are important in presupernova models. Therefore, knowledge of the experimental GT distributions should be elaborated and theoretical efforts should be made to reproduce them.

We present our formalism for the calculation of GT strength distribution and electron capture rates briefly in Section II. The GT strength distribution is presented and compared with measurements and other calculations in Section III. In this section we also present, for the first time, the microscopic calculation of GT strength distribution from the first two excited states of $^{50}$V. The pn-QRPA electron capture rates for $^{50}$V are presented in Section IV. Here we also compare our rate calculations with those of FFN and large-scale shell model calculations. Section V finally summarizes this work.

## II. METHOD OF CALCULATION

The QRPA theory is an efficient way to generate GT strength distributions. These strength distributions constitute a primary and non-trivial contribution to the electron capture rates among iron-regime nuclides. We used the pn-QRPA theory to calculate the strength functions and the associated electron capture rates for $^{50}$V. The reliability of the pn-QRPA calculations was discussed in detail by Nabi and Klapdor [13]. There the authors compared the measured data (half lives and B(GT) strength) of thousands of nuclides with the pn-QRPA calculations and got fairly good comparison. We incorporated experimental data wherever available to further strengthen the reliability of our calculated rates. The calculated excitation energies (along with their log $ft$ values) were replaced with an experimental one when they were within 0.5 MeV of each other. Missing measured states were inserted and inverse and mirror transitions were also taken into account. We did not replace the theoretical levels with the experimental ones beyond the excitation energy for which experimental compilations have no definite spin and/or parity.

The pn-QRPA theory was used with a separable interaction which granted us the liberty of performing the calculations in a much larger single-particle basis than a general interaction. We performed the pn-QRPA calculations using seven major harmonic oscillator shells. The Hamiltonian for our calculations was of the form

$$H^{QRPA} = H^{sp} + V^{pair} + V^{ph}_{GT} + V^{pp}_{GT},$$

where $H^{sp}$ is the single-particle Hamiltonian, $V^{pair}$ is the pairing force, $V_{GT}^{ph}$ is the particle-hole (ph) Gamow-Teller force, and $V_{GT}^{pp}$ is the particle-particle (pp) Gamow-Teller force. Single-particle energies and wave functions were calculated in the Nilsson model [14], which takes into account nuclear deformations. The proton-neutron residual interactions occur in particle-hole and particle-particle interaction forms. These interactions were characterized by two interaction constants $\chi$ and $\kappa$, respectively. In this calculation, we took the values of $\chi = 0.1000$ MeV and $\kappa = 0.1000$ MeV for $^{50}$V (for further details of choice of these parameters we refer the reader to [15]). The electron capture rates from the $i^{th}$ state of the parent to the $j^{th}$ state of the daughter nucleus is given by

$$\lambda^{ec}_{ij} = \left[\frac{\ln 2}{D}\right]\left[f_{ij}(T,\rho,E_f)\right]\left[B(F)_{ij} + \left(g_A/g_V\right)^2 B(GT)_{ij}\right]$$

We took the value of D=6295s [16] and the ratio of the axial vector to the vector coupling constant as -1.254 [17]. Since then these values have changed a little bit but did not lead to any appreciable change in our capture rate calculations. $B_{ij}$'s are the sum of reduced transition probabilities of the Fermi B(F) and GT transitions B(GT). Details of the reduced transition probabilities can be found in [13, 18, and 19]. For details regarding the QRPA wave functions we refer to [20]. The phase space integral ($f_{ij}$) is an integral over total energy and for electron capture rates it is given by

$$f_{ij} = \int_{w_l}^{\infty} w\sqrt{w^2-1}(w_m+w)^2 F(+Z,w) G_- dw,$$

In the above equation, $w$ is the total energy of the electron including its rest mass, $w_l$ is the total capture threshold energy (rest + kinetic) for electron capture. $G_-$ is the Fermi-Dirac distribution function for electrons. In the above equation, $F(+z,w)$ are the so-called Fermi functions and are calculated according to the procedure adopted by Gove and Martin [21]. The total electron capture rate per unit time per nucleus is then calculated using

$$\lambda^{ec} = \sum_{ij} P_i \lambda^{ec}_{ij}.$$

The summation over all initial and final states is carried out until satisfactory convergence in the rate calculations is achieved. Here $P_i$ is the probability of occupation of parent excited states and follows the normal Boltzmann distribution.

### III. COMPARISONS OF GT STRENGTH DISTRIBUTIONS

Gamow-Teller (GT) transitions are the dominant excitation mode for electron captures during the presupernova evolution. The GT+ strength distributions for the electron capture rates on $^{50}$V have been calculated using the pn-QRPA theory. Quenching of the GT strength is taken into account and a standard quenching factor of 0.77 ([22], and references therein) is used. We calculated microscopically the GT strength for 114 excited states of $^{50}$V up to excitation energies in the vicinity of 9 MeV. For each excited state of parent $^{50}$V we considered 100 states in the daughter $^{50}$Ti up to energies around 30 MeV. We employed a luxurious model space of $7\hbar\omega$ in our calculations as mentioned in previous section.

Our calculated GT strength distribution for the ground state of odd-odd nucleus $^{50}$V is shown in Fig. 1, while the (d,$^2$He) experiment and shell model results are shown in Fig.2 and Fig.3, respectively (both Figures 2 & 3 are adapted from Ref.[23]). The peak at energy 9.1 MeV (Fig. 1) towers up to a value of 1.5 on the B(GT+) axis. Because of the high level density and the size of the statistical error, a comparison of the individual peaks can not be made. Bäumer and collaborators [23] restricted the extraction of GT strength to $E_x \leq 12.2$ MeV due to the statistical error present in their spectra. Taking into account the transition up to 12.2 MeV, our calculation shows that the centroid of the GT strength resides at energy 8.8 MeV, which is in total agreement with the experimental value [23]. The GT centroid of the shell model calculation was calculated to be 8.5 MeV [23]. We placed our centroid at a slightly higher energy as compared to the shell model results. Table I presents the measured and calculated (QRPA and shell model) values of the centroid and total GT strength for the ground state of $^{50}$V up to 12.2 MeV in daughter $^{50}$Ti. The total GT strength calculated by the QRPA model lies very close within the upper limit of the measured value. This is to be contrasted with the shell model strength which coincides with the lower limit of the experimental value.

Large-scale shell model calculations usually apply the so-called Brink's hypothesis as an approximation to calculate GT strength distributions from excited states. Brink's hypothesis states that GT strength distribution on excited states is identical to that from the ground state, shifted only by the excitation energy of the state. We on the other hand, present for the first time, the microscopic calculations of these strength distributions for the excited states of $^{50}$V. Fig. 4 shows the GT strength distribution for the ground and first two excited states (at 0.23 and 0.32 MeV, respectively) of $^{50}$V. At high temperatures and densities pertinent to supernova conditions these excited states have a finite probability of occupation and a microscopic calculation of strength function from these states is desirable. The GT centroids for the first and second excited states are around 8.8 MeV and 9.0 MeV, respectively, in daughter $^{50}$Ti (up to 12.2 MeV). The corresponding total GT strengths for 1$^{st}$ and 2$^{nd}$ excited states are 7.9 and 11.2, respectively. The extracted total GT strengths from excited states also suggest that the Brink's hypothesis may not be a good approximation to use.

## IV. ELECTRON CAPTURE RATES

The electron capture rates were calculated as per the formalism presented in Section II. In this section we present these calculations and compare them with the FFN and large-scale shell model results. We present our electron capture rates at an abbreviated temperature-density scale in Table II. A complete set of calculated rates may be requested as ASCII files from the corresponding author.

Fig.5 depicts the comparison of our calculated rates with the FFN rates and that calculated using shell model. The comparison is shown for three values of densities ($\log\rho Y_e = 3, 7, 11$ g/cm$^3$). Densities are measured in units of g/cm$^3$ while $T_9$ measures the temperature in $10^9$ K. The scales are appropriate for both types of supernovae (Type-Ia and Type-II). The first two panels show the electron capture rates calculated by pn-QRPA (this work) and shell model [7], respectively. At low temperatures ($T_9 \sim 1$) and densities ($\log\rho Y_e < 7$ g/cm$^3$) the QRPA rates are smaller than the shell model rates. At ($\log\rho Y_e = 3$ g/cm$^3$), our rates are suppressed by roughly three orders of magnitude. Whereas at higher densities ($\log\rho Y_e \sim 7$ g/cm$^3$) our rates are suppressed by one order of magnitude compared to shell model results. Simulators should take note of our suppressed electron capture rates during the presupernova stage. We recall that we placed the centroid of our strength at slightly higher value of energy in the daughter. As the temperature of the stellar core increases ($T_9 \sim 3$) we are in good comparison with the shell model rates. At high temperatures ($T_9 \sim 10$) and densities ($\log\rho Y_e \sim 11$ g/cm$^3$) our rates are slightly enhanced compared to shell model (roughly by a factor of five). In the domain of high density region the Fermi energies of electrons are high enough and the electron capture rates are sensitive to the total GT strength rather than its distribution details [24]. Our calculated total GT strength is larger as compared to shell model strength (see Table I) and results in the enhancement of our rates in the high density region. We also note here that we did not employ the so-called Brink's hypothesis in our calculations as usually employed by large-scale shell model calculation.

The FFN electron capture rates are depicted in the right panel of Fig.5. Comparing the QRPA rates with FFN calculations we found a similar trend at presupernova densities and temperatures (as was the case with the shell model rates). At high densities ($\log\rho Y_e \sim 11$ g/cm$^3$) our rates are in reasonable agreement with the FFN rates.

## V. SUMMARY AND CONCLUSIONS

The Gamow-Teller strength distribution in $^{50}$V has been calculated using the pn-QRPA theory. Our results are consistent with $^{50}$V(d,$^2$He)$^{50}$Ti experiment [23]. Our total GT strength of 2.51 is higher than the shell model result of 1.42. The differences in the GT strength distributions using various model calculations were highlighted. GT strength distributions from excited states were also calculated using the pn-QRPA model and presented. The associated electron capture rates in stellar environment calculated using various models were also compared.

Type-Ia SNe contribute to about half of the abundance of the Fe-group nuclides in the galactic evolution [25]. It is desired that the simulation calculations do not overproduce these nuclides compared to their relative solar abundances. Recently Brachwitz et al. [8] did the model calculations for Type-Ia supernova by employing separately FFN and shell model rates set. The calculations resulted in the overproduction of $^{50}$Ti by using the electron capture rates of FFN. The outcome of the simulation calculations, however was much improved when (the relatively suppressed) rates of shell model were employed. It might be interesting to study how the composition of the ejecta using presently reported QRPA rates compare with the observed abundances by performing similar calculations. We are currently working on calculations of weak rates of other iron regime nuclides in large model space (up to $7\hbar\omega$) which are believed to play pivotal role in the neutronization of the nucleosynthesis ejecta and hope to report soon on our results.

**ACKNOWLEDGEMENT**

This work is partially supported by the ICTP (Italy) through the OEA-project-Prj-16.

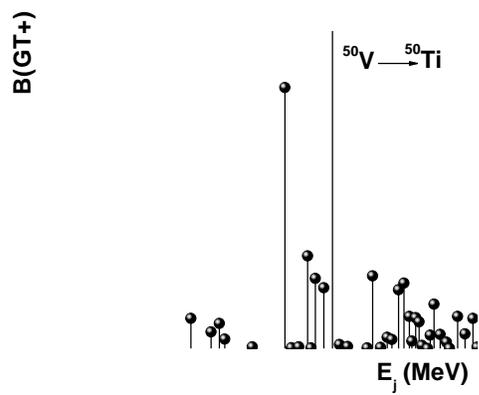

FIG.1. Gamow-Teller strength distribution for the ground state of $^{50}$V. The energy scales refers to the excitation energy in the daughter $^{50}$Ti.

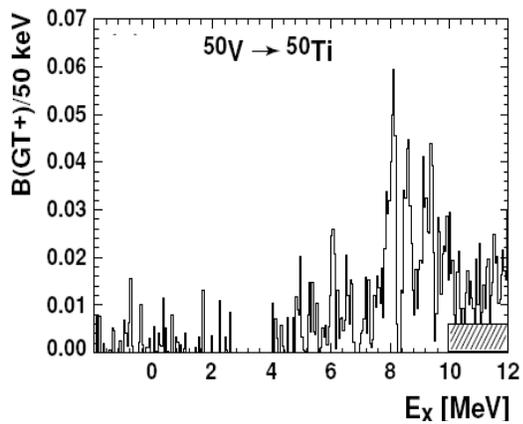

FIG.2. Experimental GT+ strength distribution for $^{50}$V $\rightarrow$ $^{50}$Ti measured through $^{50}$V(d,$^{2}$He)$^{50}$Ti. The hatched area marks the contribution from non-$\Delta L = 0$ transition strength. (The figure is reproduced from Ref [23]).

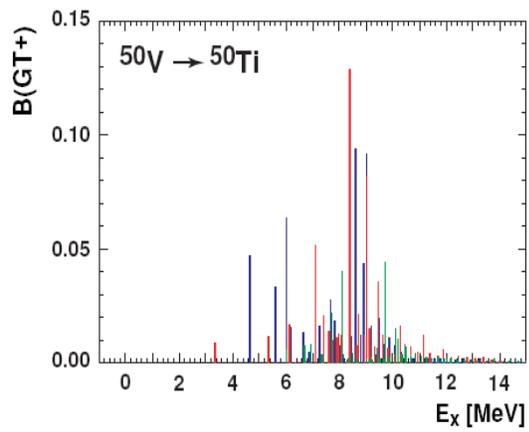

FIG.3. Large scale shell model calculations of GT strength distributions for the ground state of $^{50}$V.(The figure is reproduced from Ref [23]).

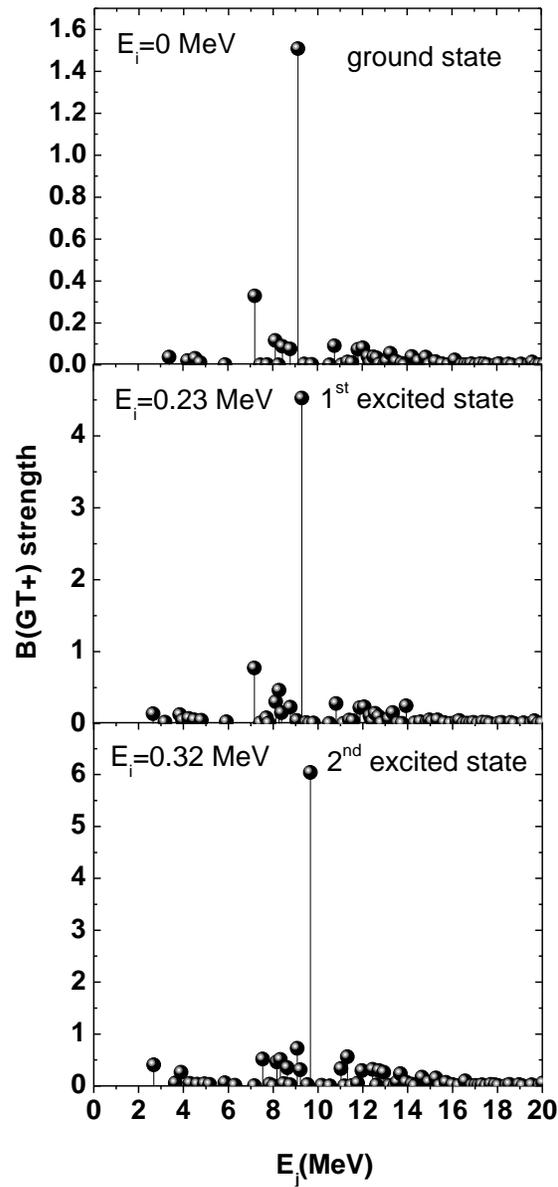

FIG. 4. Gamow-Teller (GT) strength distributions for $^{50}$V. The top, middle and bottom panels show our calculated GT strength distributions for the ground state, 1$^{st}$ and 2$^{nd}$ excited states, respectively. Ei (Ej) represents the parent (daughter) states (in MeV).

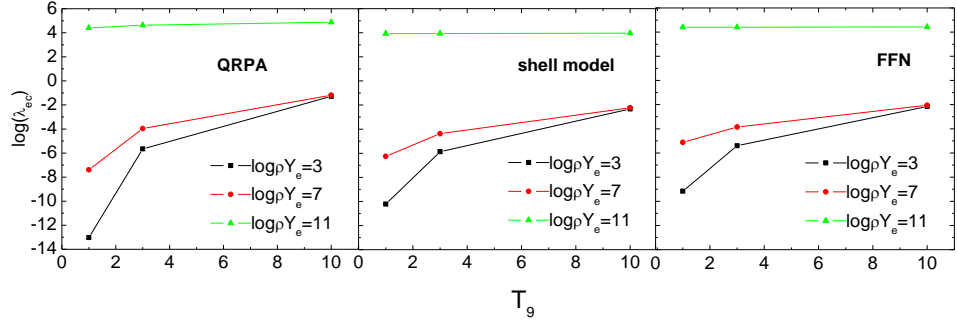

FIG. 5. QRPA electron capture rates for $^{50}$V as a function of temperature for different selected densities (left panel). The middle and the right panels show corresponding rates using the large-scale shell model [7] and FFN [2] calculations, respectively. For units see text.

TABLE I: Comparison of GT centroids and total GT strengths (up to 12.2 MeV)

| Models | GT centroid | $\sum S(GT+)$ |
|---|---|---|
| QRPA | 8.8 MeV | 2.51 |
| Experiment | 8.8±0.4 MeV | 1.9±0.5 |
| Shell model | 8.5 MeV | 1.42 |

TABLE II: Calculated electron capture rates for $^{50}$V ($^{50}$V $\longrightarrow$ $^{50}$Ti) for different selected densities and temperatures in stellar matter. $\rho Y_e$ has units of g/cm$^3$, where $\rho$ is the baryon density and $Y_e$ is the ratio of the electron number to the baryon number. Temperatures (T$_9$) are measured in $10^9$ K. The densities and calculated electron capture rates ($\lambda_{ec}$) are tabulated in log scales.

| log $\rho Y_e$ | T$_9$ | log $\lambda_{ec}$ |
|---|---|---|
| 1.0 | 1.00 | -13.257 |
| 1.0 | 3.00 | -5.651 |
| 1.0 | 10.00 | -1.289 |
| 1.0 | 30.00 | 2.472 |
| 3.0 | 1.00 | -13.029 |
| 3.0 | 3.00 | -5.650 |
| 3.0 | 10.00 | -1.288 |
| 3.0 | 30.00 | 2.473 |
| 5.0 | 1.00 | -11.176 |
| 5.0 | 3.00 | -5.591 |
| 5.0 | 10.00 | -1.287 |
| 5.0 | 30.00 | 2.473 |
| 8.0 | 1.00 | -3.428 |
| 8.0 | 3.00 | -2.111 |
| 8.0 | 10.00 | -0.559 |
| 8.0 | 30.00 | 2.508 |
| 11.0 | 1.00 | 4.409 |
| 11.0 | 3.00 | 4.631 |
| 11.0 | 10.00 | 4.871 |
| 11.0 | 30.00 | 5.539 |